\documentclass{article}

\usepackage{arxiv}

\usepackage[utf8]{inputenc} 
\usepackage[T1]{fontenc}    
\usepackage{hyperref}       
\usepackage{url}            
\usepackage{booktabs}       
\usepackage{amsfonts}       
\usepackage{nicefrac}       
\usepackage{microtype}      

\usepackage{graphicx}
\usepackage{doi}
\usepackage {lineno}

\usepackage{mathtools}
\usepackage{amssymb} 
\usepackage{amsmath}
\usepackage{caption}
\usepackage{subcaption}
\usepackage[ruled,vlined,linesnumbered ]{algorithm2e}

\newcommand \Until      {\mathbin{\mathcal{U}\kern-.1em}}
\newcommand \Iff        {\mathbin{\leftrightarrow}}

\title{Towards formalization and monitoring of microscopic traffic parameters using temporal logic}

\author{
\textbf{Mariam Nour, M.Sc}\\
  PhD Student\\	
  Department of Civil, Environmental, and Construction Engineering\\
  University of Central Florida\\
  12800 Pegasus Drive, Orlando, FL 32816\\
  
	\texttt{mariam.nour@knights.ucf.edu} \\
	\And
	\textbf{Mohamed H. Zaki, Ph.D}\\
  Assistant Professor\\
  Department of Civil, Environmental, and Construction Engineering\\
  University of Central Florida\\
  12800 Pegasus Drive, Orlando, FL 32816\\
	\texttt{mzaki@ucf.edu} \\
}

\hypersetup{
pdftitle={A template for the arxiv style},
pdfsubject={q-bio.NC, q-bio.QM},
pdfauthor={David S.~Hippocampus, Elias D.~Striatum},
pdfkeywords={First keyword, Second keyword, More},
}

\begin{document}
\maketitle

\begin{abstract}
Smart cities are revolutionizing the transportation infrastructure by the integration of technology.
However, ensuring that various transportation system components are operating as expected and
in a safe manner is a great challenge. 
In this work, we propose the use of formal methods as a means to specify and
reason about the traffic network’s complex properties. Formal methods provide a flexible tool to
define the safe operation of the traffic network by capturing non-conforming behavior, exploring
various possible states of the traffic scene and detecting any inconsistencies within it. Hence,
we develop specification-based monitoring for the analysis of traffic networks using the formal
language, Signal Temporal Logic. We develop monitors that identify safety-related behavior such
as conforming to speed limits and maintaining appropriate headway.
 The framework is tested using a calibrated micro-simulated highway scenario and offline specification-based monitoring is applied to individual vehicle trajectories to understand whether they violate or satisfy
the defined safety specifications. Statistical analysis of the outputs show that our approach can differentiate
violating from conforming vehicle trajectories based on the defined specifications. This
work can be utilized by traffic management centers to study the traffic stream properties, identify
possible hazards, and provide valuable feedback for automating the traffic monitoring systems.
\end{abstract}


\section{Introduction}
Smart mobility is revolutionizing the way modern transportation network is envisioned. It holds the promise of a more sustainable, safer, and efficient future for our commutes. However, ensuring that the various components of the urban road networks, such as traffic systems, are operating optimally and safely is not a trivial task. As a result, growing research efforts on traffic monitoring systems are being performed to design and implement novel approaches for safety applications. Traffic monitoring systems collect and analyze necessary data for the proper operation, evaluation, and management of the overall transportation infrastructure \cite{de_Souza_Brennand_Yokoyama_Donato_Madeira_Villas_2017},\cite{khan2017unmanned},\cite{lu2018using}. The collected data are invaluable in understanding the traffic dynamics, identifying issues, and bringing forth practical solutions. Current approaches to traffic monitoring systems depend on traffic detector sensors, Vehicular Adhoc Networks (VANETs) \cite{sandt2021estimating}, or video processing for data collection \cite{Jain_Saini_Mittal_2019}, \cite{mosier2018vehicle}, \cite{sandt2021estimating}. Traffic detector sensors such as roadside sensors, Bluetooth detection, or inductive loops usually capture basic traffic information such as speed, headway, or counts. VANETs, while not designed as sensors but rather as a way of communication between vehicles, can be used to transmit more detailed traffic-related information such as position, distance to surrounding vehicles and safety warnings. Video data also can capture many of the above information in addition to detailed trajectories information. With advances in machine learning, a detailed interpretation of the traffic scenes is now also possible. However, none of the above technologies can provide high-level reasoning about the state of the traffic. Commonly, this task is achieved through manual visualization of the captured traffic data. Therefore, this paper proposes the use of formal methods in the problem of traffic monitoring.

Formal verification methods were initially developed to offer a rigorous approach for analyzing computer systems, exploring the possible states that the system can visit, and detecting any inconsistencies in its operation. Inconsistencies can be in the form of incorrect implementation of the system requirements, contradicting rules that govern the system operations or even bugs in implementing the system. Analyzing systems for these purposes can be time consuming and complicated. Therefore, mathematical functions have been proposed for the task of mapping these complex behaviors into logical values which would indicate whether the performance of these systems are as expected.
Formal verification requires precise specification language to describe the system requirements and a model to describe the system's different states. Variants of formal languages exist, as discussed in the literature review section of this paper. Many of those approaches have been applied successfully to analyze and monitor complex systems such as cyber-physical systems \cite{Platzer_verif_CPS_trans} and recently to transportation systems \cite{formal_methods_ITS}, \cite{Ma_runtime}, \cite{Coogan_Arcak_Belta_2017} \cite{Rashid_Umair_Hasan_Zaki_2020}. The use of formal methods in transportation applications can provide a means to specify and reason about the traffic network's complex properties in a rigorous manner. This new way of traffic monitoring and verification can uncover potential shortcomings in the network operators that might not be detected using the current state-of-the-art methods.

The primary goal of this paper is to adopt specification-based monitoring for analyzing vehicle trajectories. Our approach focuses on identifying safety-related behavior in the traffic stream, such as conforming to speed limits and maintaining appropriate car-following behavior. 

In this work, the main contribution is the formalization of microscopic traffic behavior using
Signal Temporal Logic. These formalized specifications can be used to analyze different complex patterns in traffic. A calibrated highway scenario is implemented using the micro-simulator, SUMO, and the developed specifications are then applied to the extracted vehicle trajectories. A data dissemination algorithm was implemented using inter-vehicular communication. The purpose of this algorithm is to provide another case study for the application of the proposed approach. The value of this framework can be realized by traffic management centers for the purposes of understanding the traffic stream properties, define hazards and produce solutions. The proposed specifications can be the base of synthesis rule that governs interaction between different road users, autonomous and non-autonomous vehicles. This will eventually lead to safe deployment and operation of novel means of transportation

\section{Literature Review}

\subsection{Formal Methods in Intelligent Transportation Systems}
The application of formal methods in the transportation field is gaining traction. The majority of the research in this area is focused on ensuring the overall safety of the transportation systems. In this section, the use of formal methods in the field of verifying the safety of autonomous vehicles are discussed and their application in transportation systems as a whole. 


Arising from the need to complement classical modelling and simulation methods, researchers are shifting towards the use of formal methods for defining safety properties of autonomous vehicles. It has been proven that the use of statistical data-driven approaches for this problem is not feasible, as they require huge amounts of data to reach satisfactory results that can be accepted by the public \cite{formal_model_vehicles}. Therefor, formal methods are often used in the verification of intra-vehicle systems for autonomous vehicles, to ensure the correct operation of the most critical components such as decision making and software control modules \cite{ter_Beek_Gnesi_Knapp_2018}.

Various studies discuss the different aspects of autonomous vehicle safety such as; collision avoidance, lane change maneuvers and adaptive cruise control.
In \cite{formal_model_vehicles}, Shalev-Shwartz et al. aim to exhaustively and formally define a safety model for the nominal safe operation of self-driving vehicles. This is achieved by defining different properties and scenarios that the AV might find itself in and the proper response to each scenario. They also define both the driving policy as well as the sensing system that would achieve a safe and comfortable experience.  Mao et al. \cite{Mao_Chen_2012} defined temporal specifications for the safe operation of Cooperative Adaptive Cruise Control (CACC) and the results showed that their approach can capture specification violation correctly. 


On the other hand, there's research concerned with the development of formal models for the improvement of the traffic system as a whole. For example, the work by Mitsch et al. \cite{Platzer_2012_freeway} was one of the early attempts to utilize formal verification tools in the modelling of freeway dynamics. The objective was to ensure the system correctly calculates the appropriate speed limit and communicates this information to vehicles in certain regions of interest, this would ultimately create an online variable speed limit system. Differential dynamic logic was used for the formulation and verification of the system specifications.

Coogan and Arcak \cite{LTL_Coogan_2014} propsoed Linear Temporal Logic (LTL) specifications to define a ramp metering control strategy by using a Cell Transmission Model for the traffic system. CTM defines the system as a series of flows being controlled by upstream demands and downstream supplies. Multiple LTL formulae were proposed to ensure the design objectives of the controller are met, such as  safety, reachability and liveness.
Building further on this work, Kim et al. \cite{LTL_Kim} used the same concept of upstream demands and downstream supplies to build traffic controllers in sub-networks that would guarantee the soundness of the global network controller. This was achieved by creating supply and demand contracts between the sub-networks using LTL specifications. The concept of contracts was also explored by Müller et al. \cite{Muller_Mitsch_Platzer_2015} to show that proving the safety of one component, ensures the safety of the overall traffic system. A directed graph was developed to model the signalized intersection and differential dynamic logic is used for the formal verification of the whole system.
Other work by Coogan et al.\cite{LTL_Coogan_2016}, \cite{LTL_Coogan_2015} formalized controllers for signalized intersections using LTL to achieve more efficient traffic networks. It was argued that compared to traditional control approaches, LTL offers a rich set of expressions to address more complex traffic requirements.
For example, the latter developed LTL properties such as "a signal’s state cannot change twice in two periods", which would ensure the correct operation of the signal controller. 

Even though LTL is a powerful specification language for system description, it can be argued that it is not rich enough to capture all the complexities of the traffic dynamics. Hence, other temporal logic languages have been proposed for such purposes. The study conducted by Mehr et al. \cite{Mehr_STL} used Signal Temporal Logic (STL) to define a predictive ramp metering system for a highway in a stochastic approach. 
Rashid et al. \cite{Rashid_Umair_Hasan_Zaki_2020} focused on defining the macroscopic traffic parameters using High Order Logic specifications, while Sadraddini et al. \cite{Sadraddini_Rudan_Belta_2017} used Metric Temporal Logic (MTL) for the purpose of formalizing traffic signal control.

Ma et al. \cite{Ma_runtime} used STL specifications to create a runtime monitoring system for the improvement of safety and performance of various smart city services. These services include: environment, transportation and emergency. Ma et al. \cite{Ma_SaSTL} created a new formal verification language specifically for the use of formally define different services within a smart city. This new language extends the temporal properties from STL by adding a spatial component to it. However, none of these studies focuses on the detection and analysis of safety-related aspects in the traffic stream using formal methods. This work defines microscopic traffic parameters, with a focus on ensuring the safe bounds of these parameters.

\section{Preliminaries}

\subsection{Specification-based Monitoring}

Cyber-physical systems generate various types of data which can be in the form of time series data, wave forms or signals \cite{Bartocci_Deshmukh_Donze_Fainekos_Maler_Nickovic_Sankaranarayanan_2018}. One of the major tasks when designing and testing these systems, is to analyze their outputs to understand whether they are behaving as designed or not. This evaluation is realized through the monitoring of their temporal behavior; which can vary in length and can hold many variables and events,  hence carrying complex information. This approach is known as specification-based monitoring, which is derived from the formal verification approaches that focus on the specification of the system requirements using formal languages.

As a result, specification-based monitoring has been utilized for the purpose of ensuring that systems conform to the designer's intention. This is achieved by the rigorous specification of the system requirements using formal languages. Many formal languages have been proposed for system specifications that use temporal logic or regular expressions. However, CPS have special requirements as such systems include both physical and cyber components and produce dense temporal continuous signals. Therefor, there is a need for formal languages to address these challenges. Signal Temporal Logic (STL) has emerged as one of the new prominent formal languages in this field \cite{STL_Oded}, \cite{STL_Oded_Donze_robust}. It uses predicates on numerical values as well as atomic propositions. In the next section, we discuss the syntax and applications of STL.

\subsubsection{Signal Temporal Logic}

Signal Temporal Logic (STL) is a formal specification language that is used to describe systems with real-valued, continuous signals. It was first proposed by Maler and Nickovic \cite{STL_Oded} as an extension to Metric Temporal Logic (MTL) to formalize the properties of cyber-physical systems (CPS). It defines predicates that enable reasoning over real-time properties of systems that exhibit both continuous and discrete dynamics. 

STL specifications are defined using a combination of atomic predicates, boolean and temporal operators \cite{Bartocci_Deshmukh_Donze_Fainekos_Maler_Nickovic_Sankaranarayanan_2018}. The boolean operators include negation $\lnot$, conjunctions $\vee$ and disjunctions $\wedge$, while temporal operators include (Always) $\Box$, (Until) $\Until$ and (Eventually) $\Diamond$. There are two types of temporal operators, namely future and past operators. In our work, we focus on using the future operators where the satisfaction of such operators at position $t$ is dependant on the signal's value from $t$ onward. For example, $\Box p$ is true if and only if $p$ holds every $t' > t$. It is also worth mentioning here that the temporal operators can be true either over a certain time range or infinitely. For example, $\Box_{[a,b]}p$ is true if p holds during $t' \in [t+a, t+b]$. This is particularly useful when used with signals as it allows events to occur anywhere within this interval where, unlike discrete time, it is independent of sampling points or clock ticks.  In addition to the previous specifications, the definition of instantaneous events can also be useful in defining more complicated specifications, as they enable the specification of rising or falling edges in signals.

Below is the grammar used to define STL specifications:

\begin{linenomath}
 
\begin{equation}
    \varphi ::= \mu \; | \; \lnot  \mu \; | \; \varphi \wedge \psi \ | \ \varphi \vee \psi \ | \ \Box_{[a,b]} \varphi \ | \ \varphi \Until_{[a,b]} \psi \ | \  \Diamond_{[a,b]} \varphi 
    \label{stl_operations}
\end{equation}
\end{linenomath}

The STL semantics are defined as the satisfaction relation $(\xi, t) \models \mu$, which means that the signal $\xi$ satisfies the specification $\mu$ at some point $t$ if and only if, there exists a real-valued function of the signal such that $\mu(\xi(t))$. 

Furthermore, the satisfaction of STL formulae is denoted by the below specifications:
\begin{linenomath}
\begin{equation}
 (\xi,t) \models \mu \ \Iff \  \mu(\xi(t)) > 0 
\end{equation}
\end{linenomath}

\begin{linenomath}
\begin{equation}
     (\xi,t) \models \lnot \mu \ \Iff \ \lnot ((\xi,t) \models \mu) 
\end{equation}
\end{linenomath}

\begin{linenomath}
\begin{equation}
    (\xi,t) \models \varphi \wedge \psi \Iff (\xi,t) \models \varphi \wedge (\xi,t) \models \psi
\end{equation}
\end{linenomath}

\begin{linenomath}
\begin{equation}
    (\xi,t) \models \varphi \vee \psi \Iff (\xi,t) \models \varphi \vee (\xi,t) \models \psi 
\end{equation}
\end{linenomath}

\begin{linenomath}
\begin{equation}
    (\xi,t) \models \Box_{[a,b]} \varphi  \Iff \forall t' \in [t+a,t+b], (\xi,t') \models \varphi
\end{equation}
\end{linenomath}

\begin{linenomath}
\begin{equation}
     (\xi,t) \models \Diamond_{[a,b]} \varphi \Iff \exists t' \in [t+a,t+b], (\xi,t') \models \varphi
\end{equation}
\end{linenomath}

\begin{linenomath}
\begin{equation}
    (\xi,t) \models \varphi \Until_{[a,b]} \psi \Iff \exists t' \in [t+a,t+b],  (\xi,t') \models \psi \wedge \forall t'' \in [t,t'], (\xi,t') \models \varphi
\end{equation}
\end{linenomath}

Since STL provides powerful expressive capabilities, its use in transportation is gaining traction. Many studies are using STL to describe traffic systems since such systems are becoming more integrated by combing both cyber and physical aspects \cite{Ma_runtime}, \cite{Mehr_STL}, \cite{Bartocci_Deshmukh_Donze_Fainekos_Maler_Nickovic_Sankaranarayanan_2018}, \cite{Mehr_Sadigh_Horowitz_Sastry_Seshia_2017}. We observe the vehicle trajectories as continuous real-valued signals, therefore we use STL to formalize the specifications that can then be used for the monitoring of traffic behavior.


\section{Methodology}

 \subsection{Proposed Framework}

\begin{figure}[!ht]
  \centering
       \includegraphics[width=0.9\textwidth]{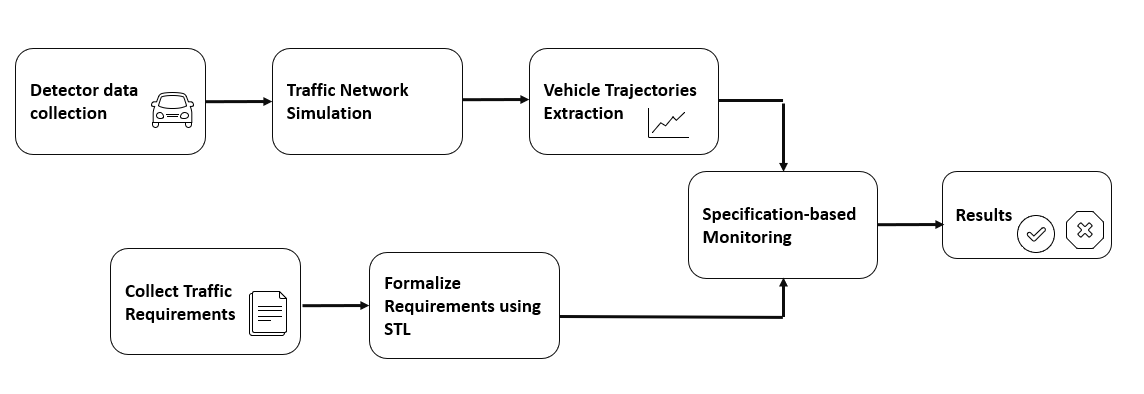}
 \caption{Methodology}
      \label{method}
\end{figure}

The methodology is split into two tasks; first is the traffic network simulation and the second is the formalization of the traffic behavior. 
As shown in figure \ref{method}, the first task begins by traffic data collection and preparation from detector data, followed by the creation of the network using the micro-simulator. The traffic network is then calibrated and the simulation is run. In this approach, a calibrated traffic network is used to test the proposed specifications \cite{gueriau2020quantifying}. The output of the traffic simulation is the set of vehicle trajectories that are used as an input to the developed specification-based monitor. As for the second task, various traffic-related rules with a focus on safety are collected; such as speed limits and safe headway. Then, these requirements are formalized using the formal verification language: Signal Temporal Logic (STL). The different formalized properties are then aggregated into a specification-based monitor that takes as an input the vehicle trajectories that were extracted from the first phase.  These specifications are developed in such a way that it's possible to unambiguously differentiate between safe and unsafe behavior in the traffic scene. For example, the monitor can understand which of the vehicles is speeding or braking in an abrupt way that would cause a disruption in the traffic stream, leading to conflict. The final step in the proposed framework is to apply the monitor to the extracted trajectories and their conformity/violation to the specifications is then studied. 

\begin{figure}[!ht]
  \centering
       \includegraphics[width=\textwidth]{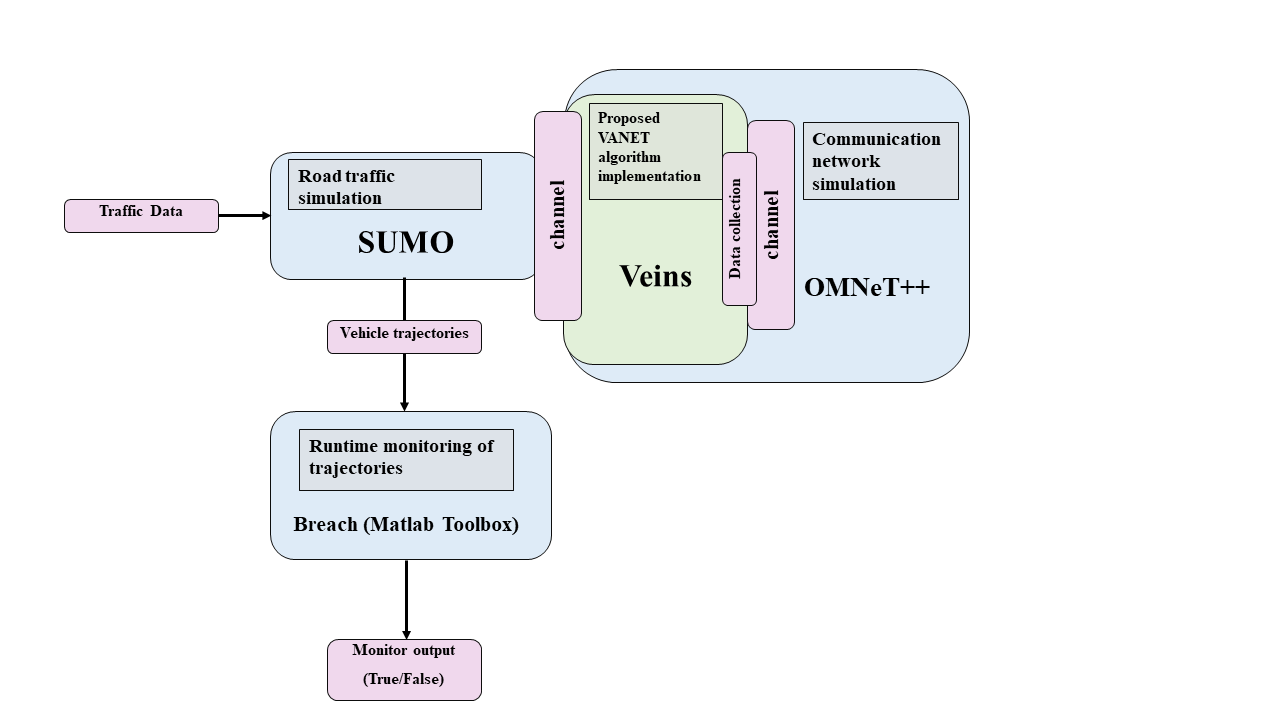}
 \caption{Proposed framework}
      \label{fig:framework}
\end{figure}

In order to implement this approach, various tools were employed as shown in figure \ref{fig:framework}. First, the obtained traffic data is used as the input to the road traffic simulator (SUMO). SUMO provides the traffic network, the vehicle types, number and behaviour through car following models. After the simulation is run and the trajectories are extracted, they are then used as an input to the runtime monitoring tool, Breach, which analyzes the conformity of these trajectories to the proposed STL specifications.
Another goal of this work is to study the impact inter-vehicle communication can have improving safety. Therefor, VEINs framework \cite{sommer2011bidirectionally} is used.  
VEINs is an open source simulator for vehicular communication. It couples both a network simulator (Omnet++) and the urban traffic network simulator (SUMO) through the TraCI protocol. The simulation starts from running an Omnet++ project that uses VEINs, where it initiates a client connection to SUMO that acts as the server. Each created vehicle is considered as a communicating node in the network as well as having the mobility defined by SUMO.

\subsection{Monitoring of Microscopic Traffic Parameters}

Inspired by the work of Bartocci \textit{et al.} \cite{Bartocci_Deshmukh_Donze_Fainekos_Maler_Nickovic_Sankaranarayanan_2018}, Signal Temporal Logic (STL) is used to formally define microscopic traffic parameters. The main focus of this work is the formal definition of speed-related parameters as well as headway parameters. This is achieved by defining these parameters in plain language and then propose STL formulae to formally describe them.

\subsubsection{Speed-related Specifications}

On a highway, vehicles must abide by certain speed-related rules. Vehicles must not accelerate in such a way that would exceed the maximum speed limit. Also, vehicles must not decelerate below the minimum speed limit as this can also pose a risk for surrounding vehicles and the traffic flow as a whole \cite{Platzer_2012_freeway}. 
Below is the formal definition of these speed requirements using STL language:

\begin{linenomath}
\begin{equation}
    \begin{split}
     \Box [(V_{min}\leq V_s \leq V_{max}) \lor  & ((V_s > (V_{max} + V_{err})) \\ &  \Longrightarrow \Diamond_{[0,t]}(V_s \leq V_{max}) \lor ((V_s < V_{min}) \\ & \Longrightarrow \Diamond_{[0,t]} (V_s \geq V_{min})) ]
    \end{split}
    \label{stl_speed}
\end{equation}
\end{linenomath}

where:

$V_{min}$: minimum speed limit

$V_{max}$: maximum speed limit

$V_s$: subject vehicle's velocity

$V_{err}$: allowed speed error 

Formula \ref{stl_speed} states that \textit{"The vehicle speed is always within the maximum/minimum allowed limits. However, if the vehicle speed increases/decreases with respect to the allowed speed limits (while allowing for some error), it eventually (during a predefined duration $t$) has to decrease/increase to be within the allowed speed limits"}. This formulation is then applied to individual vehicle trajectories to ensure that every vehicle on the highway will be conforming to the speed-related requirements. 

Another consideration when studying speed profiles of vehicles is the deceleration behavior of vehicles and studying its safety.
Jerk, which is the rate of change of acceleration, has proven to be a powerful indicator of unsafe deceleration \cite{zaki2014use}, \cite{Bagdadi_Varhelyi_2011}. Jerk can be used to differentiate between normal braking and sudden braking that is due to unsafe situations. Therefor, as a compliment measure to acceleration, jerk values can be used to accurately capture conflict situations. A threshold level of $-9.9m/s^3$ is used for the jerk as an indicator of safety-critical driving behaviour. Vehicles must also avoid decelerating abruptly as this will negatively impact the traffic flow. Hence, to define speed specification for decelerating vehicles, we need to take into consideration the safe deceleration of vehicles by including both the deceleration rate and jerk in the speed formalization. 

Below is the STL formula for speed requirements while considering safe braking behavior:
\begin{linenomath}
\begin{equation}
    \begin{split}
     \Box  [ ( A_s > -7.7)  \land (J_s > -9.9) ]
    \end{split}
    \label{stl_speed_2}
\end{equation}
\end{linenomath}

where:

$A_s$: subject vehicle acceleration

$J_s$: subject vehicle jerk

which states that \textit{"At all times, both the jerk and acceleration profiles of the vehicle must be below certain threshold, which would indicate comfortable and safe deceleration."} 

In addition to defining the deceleration behavior in general, a special case is also studied, which is defining the vehicle's speed profile while approaching an off-ramp. Vehicles intending to use an off-ramp might approach it with a higher speed than that of the off-ramp speed limit. Therefor, these vehicles need to continue deceleration until their speed matches that of the off-ramp. It's noted here that this specification is applied only to vehicle trajectories that are merging onto an off-ramp. This behavior can be defined using the following specification:

\begin{linenomath}
\begin{equation}
     \Box [(V_s \leq V_{sl})  \lor
     ( (V_s > V_{sl})  \Longrightarrow  (( A_s > -7.7) \land (J_s > -9.9)  \Until (V_s \leq V_{sl}))]
    \label{stl_offramp}
\end{equation}
\end{linenomath}

where:

$V_s$: subject vehicle speed

$V_{sl}$: maximum speed limit on an off-ramp

This formula states that \textit{"The speed of the vehicle has to always be less than the speed limit, however, if it increases to be more than the speed limit, then the vehicle must continue to decelerate comfortably until its speed is less than or equal to the maximum speed limit allowed."}


\subsubsection{Headway Specifications}

Headway is one of the fundamental microscopic traffic parameters, where it describes the temporal relation between consecutive vehicles. It can be defined as the time difference between a leader vehicle arrival at a designated point and the following vehicle reaching the same point. Headway estimation has been the focus of a lot of research as it can be implemented in various applications in traffic safety. One of these applications is the implementation of advanced driver-assistance systems (ADAS), which can alert the driver about hazards on the road. Headway estimation can be used to calculate when the driver should be warned about possible conflicts \cite{zhu2020impact}.

According to \cite{Rizaldi_Immler_Althoff_2016} , it is recommended to keep a minimum following distance of 4 seconds during normal weather and traffic condition at all times. Therefore, we use the below STL formula to define the headway specification:

\begin{linenomath}
\begin{equation}
     \Box[ (h \geq 4) \land ((h < 4) \Longrightarrow (\Diamond _{[0,t]} (h \geq 4)))]
    \label{stl_hw}
\end{equation}
\end{linenomath}

where:

$h$: headway between every two vehicles following one another

Equation \ref{stl_hw} states \textit{ the headway should always be greater than or equal to 4 seconds. However, we allow for the headway to fall below this value but only for some time $t$, if it doesn't increase back to 4 seconds or more, then the specification is violated.} This definition ensures that the individual vehicles are always abiding by the headway specifications for a safe operation of the overall traffic flow.

\subsection{Data Dissemination VANET Algorithm for Traffic Safety Improvement}

In order to enhance the overall traffic safety, we propose the use of Vehicular Adhoc Networks for the dissemination of safety-critical information to vehicles. The appropriate accelerations is calculated for each vehicle to maintain safe distance to its preceding vehicle. In this section, a case scenario is demonstrated where using the proposed STL monitors can be applied to understand the traffic behavior. An example of the headway trajectories before and after applying the communication algorithm is explained and the comparison is done through using the proposed monitors.

In general, a follower vehicle is tasked with maintaining an appropriate headway to its leader vehicle at all times by regularly actuating its acceleration \cite{Fernandes_Nunes_2012}. The calculated acceleration is based on: the follower vehicle's velocity, the leader's relative velocity and the spacing between both vehicles \cite{phd_acceleration}. Vehicular communication can be utilized as a convenient method in conveying such information among the road-users. 

The task of each follower vehicle is to maintain a certain "desired minimum gap", which is defined by the below equation:

 \begin{equation}
     s(v_\alpha, \Delta v_\alpha) = s_0 + vT + \frac{v \Delta v}{2 \sqrt{ab}}
     \label{safe_distance}
 \end{equation}

where $s_0$ is the jam distance, reasonable ranges are from $1m-5m$ so a value of $2m$ is used in our approach.

Equation (\ref{safe_distance}) is the desired minimum gap between two vehicles according to the Intelligent Driver Model (IDM) \cite{phd_acceleration}. The term $vT$ plays an important role in non-stationary scenarios, as it ensures a constant time gap regardless of the speed. It is also noticed that the desired minimum gap is directly proportional to the speed of the vehicles, which means that it increases with the increase of the speed. If maintained, this formulation of the desired gap guarantees collision-free behavior. 

Building further on this concept, equation \ref{acc_eq} defines the IDM acceleration function which takes advantage of the safety properties of the desired minimum gap equation (\ref{safe_distance}). Each follower vehicle utilizes the equation \ref{acc_eq} in order to calculate the appropriate acceleration to be applied in order to always maintain an appropriate inter-vehicle headway of $4s$. 

\begin{equation}
  \dot v_\alpha (s_\alpha,v_\alpha, \Delta v_\alpha) = a \left[1-\left( \frac{v_\alpha}{v_0} \right)^\delta - \left (\frac{s(v_\alpha, \Delta v_\alpha)}{s_\alpha} \right)^2 \right]  
  \label{acc_eq}
\end{equation}

where:

$v_\alpha$: velocity of the follower vehicle

$\Delta v_\alpha$:  velocity difference between follower and leader vehicles

$a$: maximum acceleration (a value of $1.4m/s^2$ is used)

$v_0$: the desired velocity 

$T$: Safety time gap (a value of $4s$ is used)

$b$: Desired deceleration (a value of $2.0 m/s^2$ is used)


In our proposed approach, we adopt a vehicular communication algorithm based on the work by Fernandes and Nunes \cite{Fernandes_Nunes_2012}. Their approach focuses on minimizing the inter-vehicle gaps in platoons for better road utilization, while our approach aims at maintaining safe time gap between vehicles at all times.
For every update cycle, vehicles broadcast their position and speed to their direct neighbors. It is assumed that each vehicle is able to calculate the distance to its leader, so if a vehicle is detected as a leader vehicle, the follower vehicle sends its position and speed information to it. Each follower vehicle maintains a record about its leader by constantly listening to beacons from the the leader that contains position and speed information. Both leader and follower vehicles update records according to vehicles entering and leaving the traffic stream through periodic beacons. Vehicles can be both a leader and a follower based on its position in the traffic stream.

The pseudo-code for leader and follower vehicles are explained in Algorithm \ref{alg_lead} and Algorithm \ref{alg_follow} consecutively.


\begin{algorithm}[]
\SetAlgoLined

\For{all \textit{vehicles}}{

Broadcast $x_{leader}$ and $v_{leader}$\;

\If{receives message from follower}{
   store follower information\;
   }

}
 \caption{Vehicle information updating algorithm: Leader}
 \label{alg_lead}
\end{algorithm}

\begin{algorithm}[]
\SetAlgoLined
 \For{all \textit{vehicles}}{

\If{receives message}{
   calculate $s_{leader}$ to leader vehicle\;
   \If{($s_{leader}/ v_{follower}) < 4s$}{
        declare vehicle as leader\;
        send $x_{follower}$ and $v_{follower}$ to leader\;
        calculate new acceleration to be applied next time-step\;
        
   }
   }

}

 \caption{Vehicle information updating algorithm: Follower}
 \label{alg_follow}
\end{algorithm}

Initially, a vehicle is unaware whether it has the role of a leader or a follower or both. Therefor, all vehicles start by broadcasting their position and speed information. The following chain of events decides the role of the vehicle. If it detects another vehicle preceding it, then it is a follower. If it receives messages that encapsulate its own ID in addition to speed and position information, then it is assumed that this information was sent by a follower and it is assigned the role of a leader. 

Algorithm (\ref{alg_lead}) defines the behavior of a vehicle with the leader role. The vehicle starts by broadcasting its position ($x_{leader}$) and speed ($v_{leader}$). Lines 3-5 indicate that the vehicle has received a message from its follower and stores the corresponding information. On the other hand, algorithm (\ref{alg_follow}) explains the behavior of the vehicle if it is assigned the follower role. Lines 2-3 indicate that the follower vehicle has received a message, that contains speed and position information. The distance to the vehicle ($s_{leader}$) is then calculated based on this information. If it is decided that the received information was from the preceding vehicle, then this vehicle is defined as the current leader. Then, the time headway to the preceding vehicle is calculated in lines 4-8 and if it is below a threshold of $4s$, then the follower calculates the appropriate acceleration to be applied according to equation (\ref{acc_eq}).


\section{Results}

In this section, a highway scenario is implemented to show the effectiveness of the proposed STL-specified traffic monitor to capture changes in the traffic flow parameters. The input to the monitor is each vehicle's trajectory and the output is the robustness measure of the overall satisfaction of each formula, as well as a plot showing where the trajectory was satisfying or violating the specification. After that, the vehicular communication algorithm is applied to the scenario, and the proposed monitor is used for further trajectory evaluation, to observe how the communication impacts the conformity of the vehicle's trajectory to the headway specification.

\subsection{M50 Highway}

\begin{figure}[!ht]
\centering
     \begin{subfigure}[b]{0.49\textwidth}
         \centering
         \includegraphics[width=0.5\textwidth]{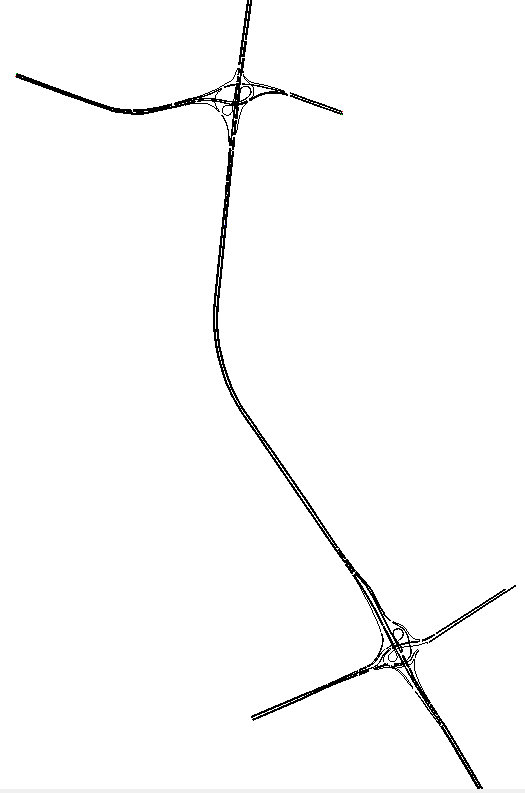}
         \caption{M50 highway in SUMO}
         \label{fig:m50sumo}
     \end{subfigure}
     \hfill
     \begin{subfigure}[b]{0.49\textwidth}
         \centering
         \includegraphics[width=0.5\textwidth]{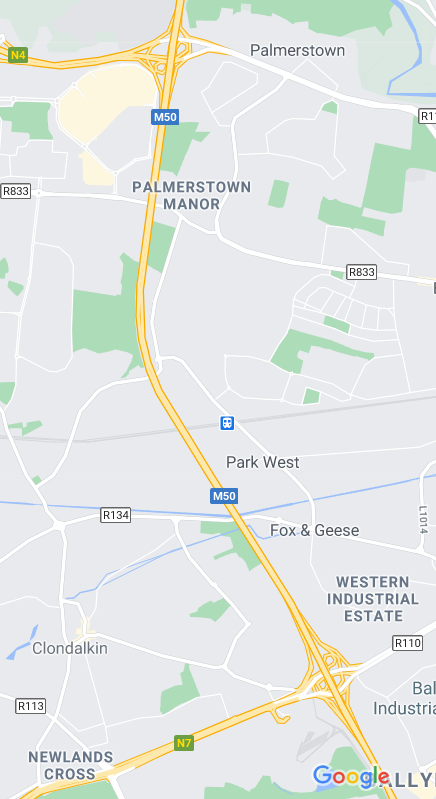}
         \caption{M50 Highway from Google maps}
         \label{fig:M50google}
     \end{subfigure}
    \caption{M50 Highway}
        \label{fig:M50}
\end{figure}

The proposed STL traffic monitor is applied to a highway scenario obtained from \cite{gueriau2020quantifying}, which was implemented and calibrated using the micro-simulator SUMO. The traffic network model simulated the M50 highway in Dublin, Ireland which is a 7KM 4-lane highway that includes two major interchanges (Figure \ref{fig:M50}). The traffic demand used in the model was generated using loop detectors from the Transport Infrastructure in Ireland, during the morning peak hours (7-8am) with 5-minute aggregated traffic flows. The simulation is run using SUMO with a time-step of $50ms$ and vehicle-specific information are extracted. This information include: speed, longitudinal and lateral dynamics. Gueriau et. al  have adjusted several simulation parameters such as headway, minimum gap values and driver imperfection factors in order to simulate the actual driving behavior as closely as possible \cite{gueriau2020quantifying}. They have used various types of vehicles such as human-driven, level 2 and level 4 connected autonomous vehicles.

\subsubsection{Speed-related Specifications Results}

The speed specification previously proposed in formula (\ref{stl_speed}) was applied to the M50 highway traffic scenario. The simulation is run for $100s$ and speed data was recorded for individual vehicles. 

\begin{figure}[!ht]
 \centering
       \includegraphics[scale=0.35,width=0.9\textwidth,height =0.5\textwidth]{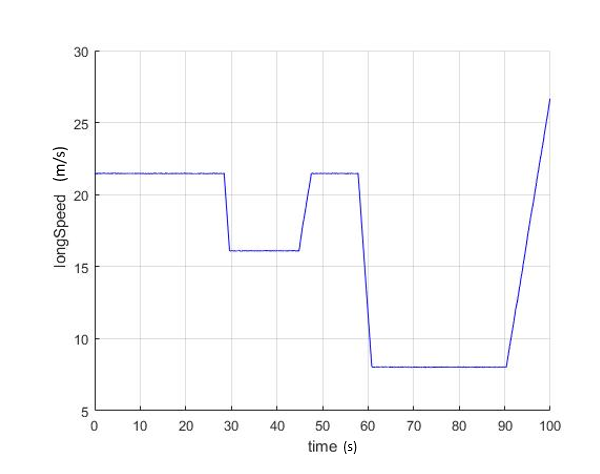}
         \caption{Speed Specification trace}
          \label{speed_trace}

 \end{figure}

In figure \ref{speed_trace}, the speed of a single vehicle is plotted across the simulation time. By observing the speed plot, it can be noticed that this vehicle speed trace doesn't conform to the speed specifications as it is below the minimum speed limit allowed on the highway. We expect that the robustness satisfaction value for this trace would be negative; which means that the input trace doesn't satisfy the specification.
The results show that the robustness satisfaction value for this trace would be negative; which means that the input trace doesn't satisfy the specification. This is confirmed by observing the robustness output signal, which indicates that the specification output is false throughout the trajectory of the vehicle. The exception to that is at $97s$, where the vehicle adjusts its speed in such a way that it conforms to the proposed speed specification. After that time, the vehicle's speed is within accepted values. The same approach can be applied automatically to all other vehicle trajectories in the highway scenario. The output would indicate whether vehicle trajectories are satisfying or violating the speed specifications on the highway.

\begin{figure}[!ht]
\centering
\includegraphics[width=\textwidth]{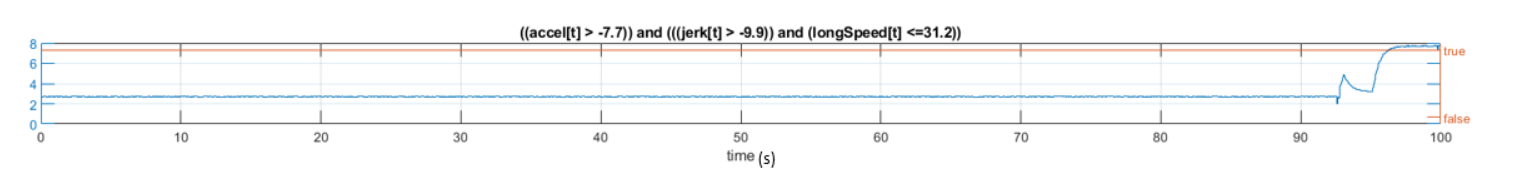}
\caption{Speed Specification for safe braking output}
\label{speed_2_out}
\end{figure}

Speed is further investigated by applying the specifications defined by the STL formula (\ref{stl_speed_2}) to individual speed trajectories. The main goal of this specification is to understand the braking behavior of vehicles in conflict situations. For safe and comfortable braking, the jerk and deceleration values of a vehicle should be bounded within values expressed by the STL formula. The specification is applied to the individual vehicle trajectories and an individual trace is used for illustrating the results. By observing trace in blue in figure \ref{speed_2_out}. It can be noted that there is a decline in speed at $95s$. To understand whether the deceleration taking place at that time was safe or not, the speed specification is applied to this trace. It is worth mentioning here that even though using high sampling rate can produce better resolution for acceleration and jerk, it can also be prone to high fluctuations and high noise \cite{Bagdadi_Varhelyi_2011}. Therefor, the acceleration and jerk profiles of the vehicle were first subjected to an exponential smoothing method using a Matlab function. This was done to smooth the data and minimize the noise.

Figure \ref{speed_2_out} depicts the output of the monitor in red, which shows that the trace is indeed conforming to the specification. It is concluded that throughout the simulation, this vehicle was exhibiting safe speed and braking behavior.

It is noted here that this way of defining speed specification with a focus on comfort using the jerk indicator can be applied in specific areas of the road segment such as off-ramps. Off-ramps usually have lower speed constraints than the main highway. Typically, vehicles on highways drive close to the highway speed limit, which would require the drivers to decelerate when approaching an off-ramp. During that time, it would be worth observing the behavior by which drivers brake in order to take the exit ramp. In this situation, it can be useful to also use our proposed speed specification \ref{stl_offramp} to study such behavior. The same vehicle trajectory studied in the previous section is used, but the difference here is that we add the condition that the vehicle is using an off-ramp. We observe how the jerk and acceleration variations in order to understand the underlying safety aspects.

\begin{figure}[!ht]
\centering
\includegraphics[width=\textwidth]{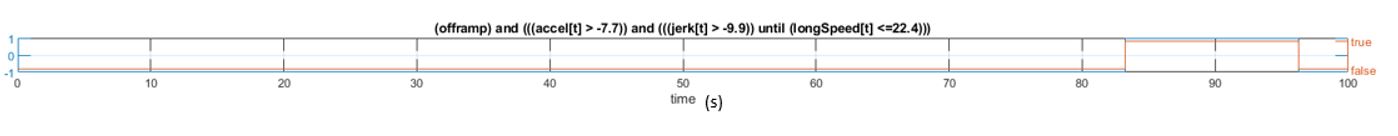}
        \caption{Speed Specification output for vehicle using off-ramp}
        \label{offramp_out}
\end{figure}

After applying the STL specification defined by formula \ref{stl_speed_2}, the output can be observed in red in figure \ref{offramp_out}. From this output, it is observed that the jerk value is greatly fluctuating during the last portion of the trace. However, the monitor registers these fluctuations as conforming behavior (red line indicates true in the last subplot) which shows that during that time, this vehicle was braking safely.


\subsubsection{Headway Specification Results}

In this section, the results of applying the headway specification \ref{stl_hw} on the M50 highway scenario is discussed. Vehicle headway information is used to test the output of the proposed monitor. It is important to note here that the trace contains some negative values, these mean that during that time, the vehicle didn't have a leader vehicle. 

\begin{figure}[!ht]
\centering
 \includegraphics[width=\textwidth]{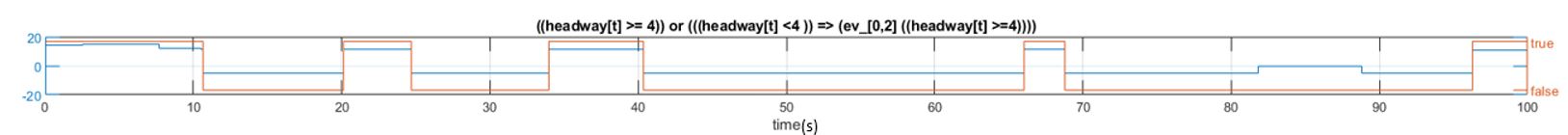}
    \caption{Headway Specification output} 
      \label{fig:headway_out}
\end{figure}

The headway trace is shown in blue in figure (\ref{fig:headway_out}). We would like to observe whether the vehicle is closely following the lead vehicle for a prolonged period of time that would result in a conflict. This behavior is defined as per STL formula (\ref{stl_hw}). The monitor output in figure (\ref{fig:headway_out}) confirms that during at all times where the vehicle has a leader, it adjusts its spacing compared to its leader within $2s$, hence the specification is satisfied. It's concluded that this vehicle is exhibiting safe car-following behavior.

\begin{figure}[!ht]
\centering
 \includegraphics[height=0.4\textwidth, width=\textwidth]{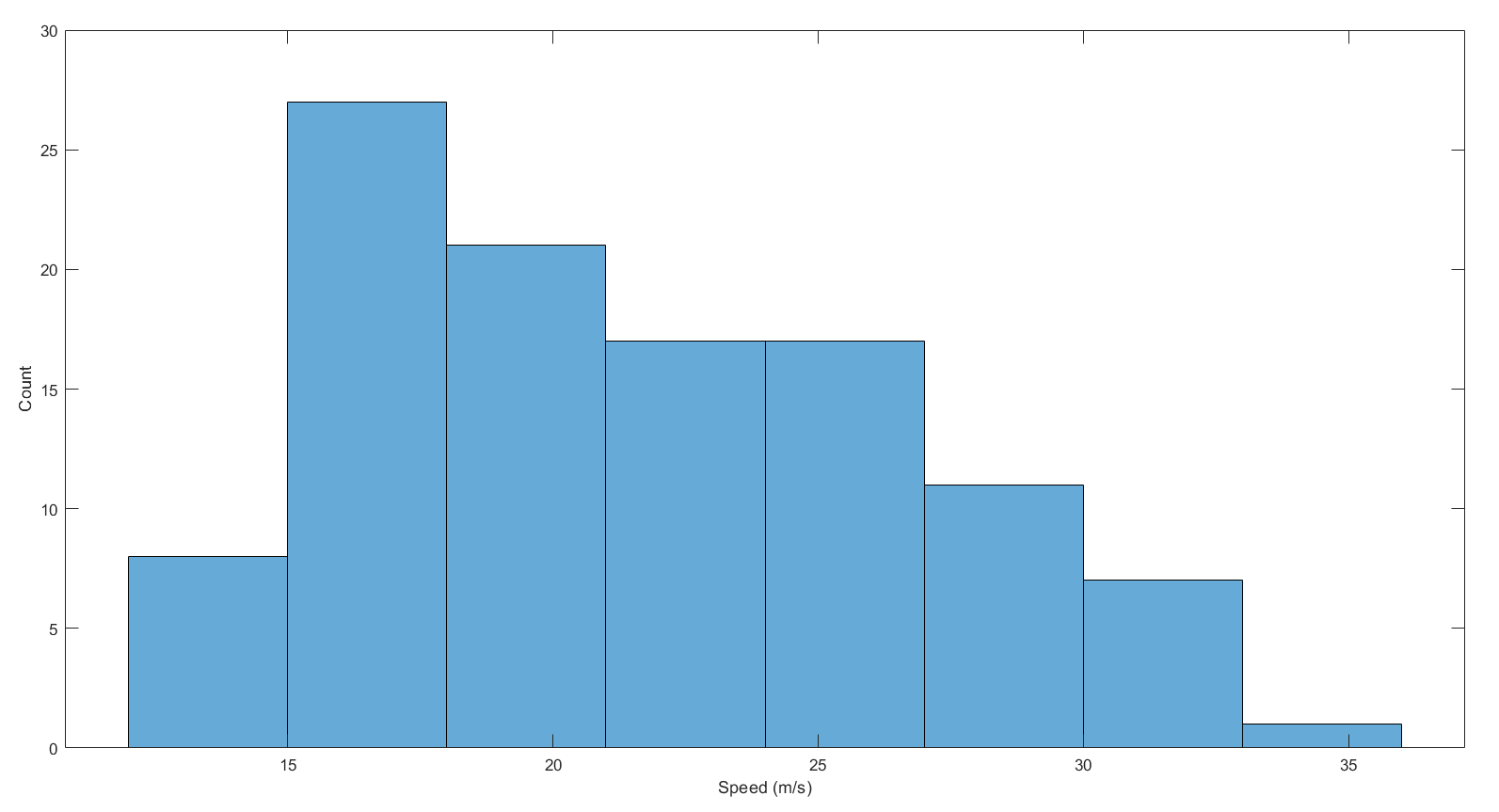}
        \caption{Speed distribution}
        \label{fig:speed_stat}
\end{figure}

To show case how the proposed monitors can be beneficial in understanding traffic characteristics, the simulation is for $150s$ and apply the speed and headway specifications to the individual vehicle trajectories. A minimum speed limit of $22.5m/s$ and a maximum speed limit of $31 m/s$ were used. Figure (\ref{fig:speed_stat}) shows the speed distribution of the scenario. The speed specifications that were introduced in equation (\ref{stl_speed}) are used to generate the following results. However, to be able to visualize the results, speed specifications are divided into two scenarios; minimum speed limit and maximum speed limit. 

The results in table \ref{table:speed_stats_min} show the minimum speed limit specification output. It's expected that vehicles with speed lower than the minimum speed limit will violate the specification. The results show that the total number of violating vehicles is 99 with a mean speed of $19.27m/s$, which is lower than the threshold set in the simulation. On the other hand, the total number of conforming trajectories is 144 with a mean speed of $23.6 m/s$.

\begin{table}[!ht]
	
	\begin{center}
		\begin{tabular}{l l l l}
			Measure & Conforming trajectories  & Violating trajectories \\\hline
		    Volume & 144 & 99 \\
            Mean Speed (m/s) & 23.66 &  19.27  \\
            Std Dev (Speed) &4.17 & 4.1 \\\hline
		\end{tabular}
	\end{center}
	\caption{Conforming vs violating trajectories to the minimum speed specification}
\label{table:speed_stats_min}
\end{table}

Next, the simulation is run to test the specification for the maximum speed limit. It's expected that vehicles with a speed higher than the maximum speed limit will violate the specification. As per results in table \ref{table:speed_stats_max}, it's found  that there are a total of 9 violating trajectories with mean speed of $27.5m/s$, on the other hand, we find a total of 232 conforming trajectories with a mean speed of $21.63m/s$.

\begin{table}[!ht]
	
	\begin{center}
		\begin{tabular}{l l l l}
			Measure & Conforming trajectories  & Violating trajectories \\\hline
		   Volume & 232 & 9 \\
            Mean Speed (m/s) & 21.63 &  27.5  \\
            Std Dev (Speed) &4.4 & 6.1 \\\hline
		\end{tabular}
	\end{center}\caption{Conforming vs violating trajectories to the maximum speed specification}
\label{table:speed_stats_max}
\end{table}

\begin{figure}[!ht]
\centering
 \includegraphics[height=0.6\textwidth, width=\textwidth]{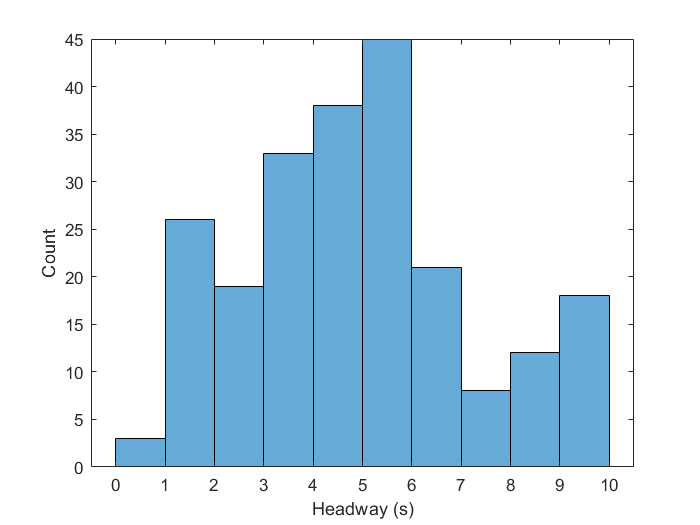}
        \caption{Headway distribution}
        \label{fig:headway_stats}
\end{figure}

In figure \ref{fig:headway_stats}, the headway distribution is shown, which was defined using formula  (\ref{stl_hw}). The  table  (\ref{table:headway_stats}) shows the mean headway of 168 conforming trajectories to be $5.17s$, while the mean headway for the 58 violating trajectories is $3.71s$. 

\begin{table}[!ht]
	
	\begin{center}
		\begin{tabular}{l l l l}
			Measure & Conforming trajectories  & Violating trajectories \\\hline
		   Volume & 168 & 58 \\
            Mean Headway (s) & 5.17 &   3.71 \\
            Std Dev (Headway) & 2.44 & 1.82 \\\hline
		\end{tabular}
	\end{center}
	\caption{Conforming vs violating trajectories to the headway specification}
\label{table:headway_stats}
\end{table}

To conclude, results from this section show that the proposed speed monitor is able to capture unsafe changes in the speed, acceleration and jerk of individual vehicle trajectories. This output can be useful in understanding the safety aspects of microscopic traffic parameters. Using the speed as a safety indicator in the proposed monitor shows when the vehicles are violating the imposed speed limits. While using indicators such as jerk help in understanding the underlying safety implications of vehicle braking behaviors. This information can be exploited by traffic management centers in studying specific highway segments where speeding can be the cause of conflicts. Another scenario where these monitors can be used is autonomous vehicles. Such vehicles require constant analysis of surrounding traffic scene, to be able to make decisions. The developed monitors can be built into autonomous vehicles for online monitoring of surrounding vehicles behavior.


\subsection{ Proposed Communication Algorithm Results}

In order to validate that the proposed communication algorithm in fact improves the safety by adjusting inter-vehicle spacing, two different simulations are run. The first case is without the application of the actuated acceleration, in this scenario, the vehicles implement the default car-following model in SUMO. The second scenario uses the proposed algorithm to exchange the safety-related information between the vehicles, hence, making it possible for follower vehicles to actuate their acceleration according to the proposed equation in \ref{acc_eq}. The simulation is run for $100s$, and a total of 110 vehicles are generated. Each vehicle has a minimum power level of $-90dBm$, transmission power of $15mW$, and a transmission range of $500m$.

\begin{figure}[!ht]
\centering
 \includegraphics[width=\textwidth]{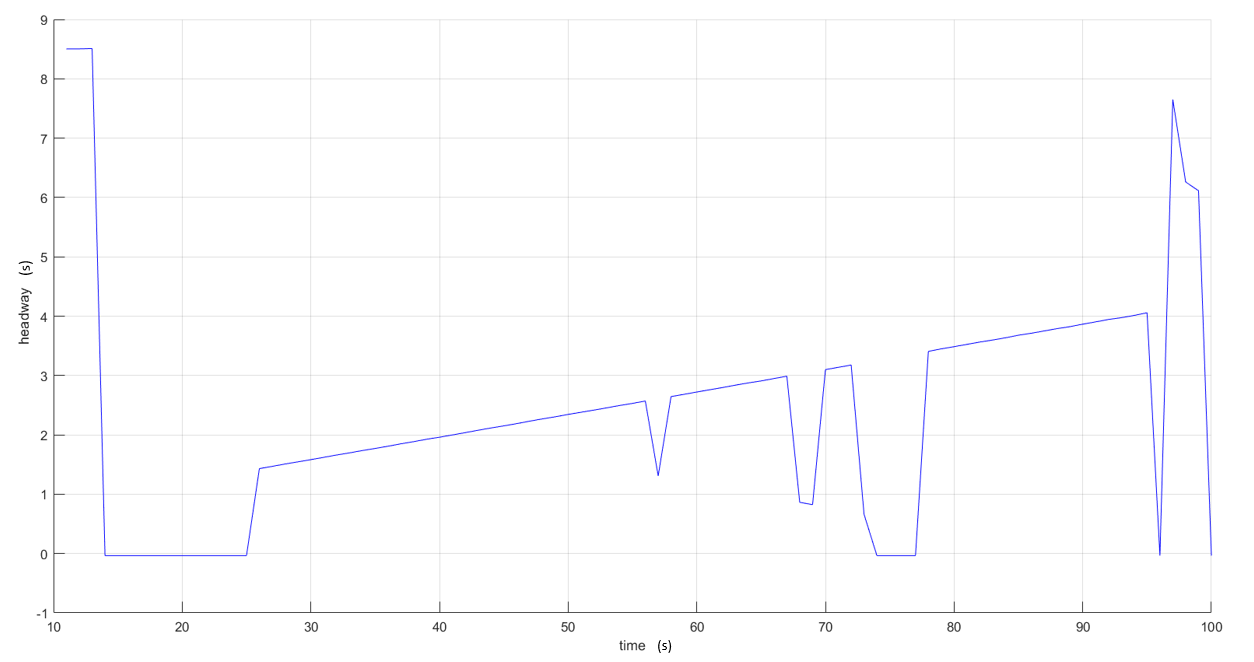}
    \caption{Headway Trace with no IVC}
    \label{noivc_trace}
\end{figure}

\begin{figure}[!ht]
\centering
\includegraphics[width=\textwidth]{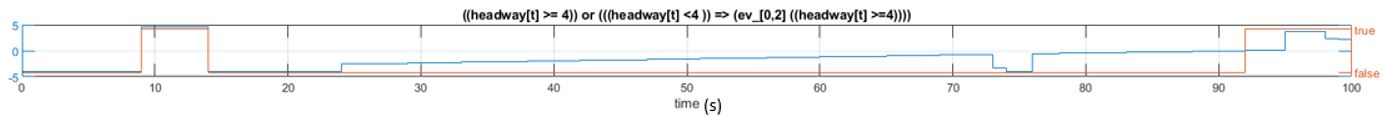}
        \caption{Headway Specification output with no IVC}
        \label{noivc_out}
\end{figure}

Next, the headway traces for the two scenarios were extracted and analyzed using the monitor defined earlier using equation \ref{stl_hw}.For the first scenario, a headway trace is shown in figure \ref{noivc_trace}, by observing the trace, it can be noticed that from around $20s$ up until $95s$ the vehicle's headway trace is below the defined threshold of $4s$. 
the monitor is applied to analyze the headway trace. As seen in figure \ref{noivc_out}, the monitor
output confirms the initial observation as it returns a false value for that same duration from $20s$ up until $95s$. 

\begin{figure}[!ht]
\centering
\includegraphics[width=\textwidth]{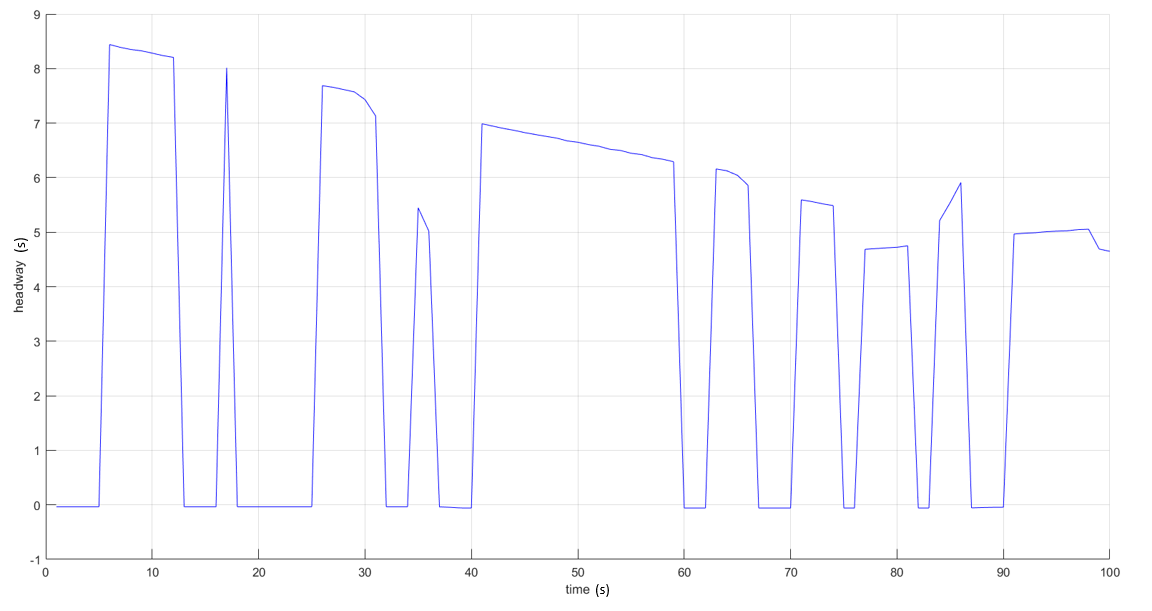}
    \caption{Headway Trace with IVC}
    \label{ivc_trace}
\end{figure}

\begin{figure}[!ht]
\centering
 \includegraphics[width=\textwidth]{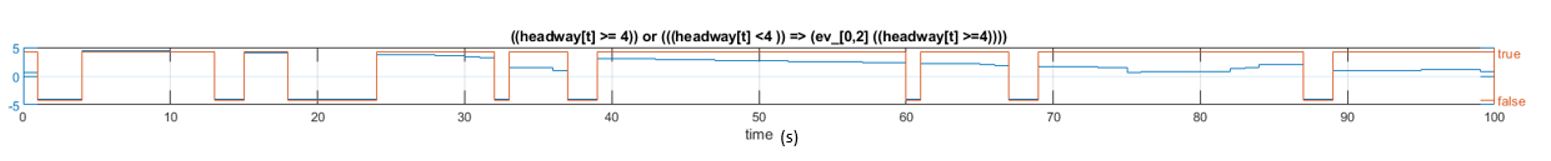}
        \caption{Headway Specification output with IVC}
        \label{ivc_out}
\end{figure}

In the second scenario,  the communication algorithm is applied to the simulation where the follower vehicles adjust their acceleration based on the disseminated information in an online fashion. 
By inspecting the headway traces from the second scenario, it can be found that the follower vehicles were complying to the headway specification. Figure \ref{ivc_trace} shows the headway trace for a vehicle, and the output trace \ref{ivc_out} shows that the trace is indeed conforming to the headway specification.

To conclude, the results of this work showed that the specification-based monitoring using STL language is able to capture various traffic stream properties. We first started by showing the speed-related specifications and how we can capture various safety aspects such as speeding, safe braking and safe exiting of a highway. The results also showed that we are able to identify safe car-following behavior based on the time headway between individual vehicles. Finally, a vehicular-communication based scenario was implemented where vehicles adjust their acceleration based on information received from leader vehicles. This enabled to showcase how the developed STL monitors can be used in practice to identify the safe headway with and without using vehicular communication.


\section{Conclusion}
In this paper, a new method for monitoring the safety of traffic behavior using formal methods was proposed. In this approach, vehicle trajectories were extracted from calibrated micro-simulation models and their conformity to safe behavior were studied by applying monitors that use formally defined specifications. The use of formal methods in traffic allows for the high-level reasoning about the traffic state. Formal methods have the ability to precisely define logical rules that would guarantee the correctness of the traces it's being applied to. As opposed to machine-learning approaches that are mainly black box models, formal methods can provide an unambiguous approach to defining systems and analyzing their behavior. This is true in the case of the traffic system. By applying the approach proposed in this paper, output from specification-based monitoring in the traffic setting can direct traffic management systems to provide solutions to specific problems and pave the way for autonomous vehicles to analyze and understand the behavior of surrounding vehicles. 

The formal language STL was used to define the different traffic-related specifications and employed a framework that consists of multiple tools for the implementation and simulation of this work. The framework consists of a traffic micro-simulator (SUMO) and a tool that defines and analyzes vehicle trajectories using formal specification-based monitoring (Breach). A highway scenario was simulated and the extracted vehicle trajectories were used to test the proposed specifications.

The results showed that the proposed approach can successfully capture different safety attributes of the traffic flow. These include vehicle trajectories conformity to speed limits, safe braking, and safe car following behavior.

One of the limitations of this work is that the proposed specifications are able to only capture temporal aspects of the traffic behavior. This can be remedied by using a more expressive specification language that would also incorporate spatial aspects. These operators would allow for describing spatial components; such as defining certain regions of interests in the traffic network.

For future work, this work can be enhanced upon to include surrogate safety measures such as time-to-collision (TTC) to study the safety of road segments. The addition of macroscopic traffic parameters monitors can provide valuable insights for traffic network safety. Other road geometries can also be used to test the synthesized traffic monitors under different conditions. Such monitors can also be extended to study more complex traffic interactions such as lane changes \cite{islam2021real}, \cite{kang2018modeling}. Finally,
online monitoring is the natural next step for the synthesis of real-time monitors. These monitors will be able to detect the specified traffic behavior and their safety in real-time. This can then be integrated within a self-contained solution that observes, analyzes and alerts decision-makers of safety hazards on the road. 

\section{Author Contributions}
The authors confirm contribution to the paper as follows: study conception and design: Mariam Nour and Mohamed Zaki; data collection: Mariam Nour; analysis and interpretation of results: Mariam Nour; draft manuscript preparation: Mariam Nour and Mohamed Zaki. All authors reviewed the results and approved the final version of the manuscript.

\bibliographystyle{IEEEtran}
\bibliography{references}  







\end{document}